%% file: main.tex
\documentclass[runningheads]{llncs}
\usepackage[T1]{fontenc}
\usepackage{graphicx}
\usepackage{multirow}
\usepackage{multicol}
\usepackage{amsfonts}
\usepackage{subfig}

\usepackage{hyperref}
\usepackage{xstring} 
\newcommand{\mystatus}{review}
\usepackage[symbol]{footmisc}

\usepackage{array}
\newcolumntype{L}[1]{>{\raggedright\let\newline\\\arraybackslash\hspace{0pt}}m{#1}}
\newcolumntype{C}[1]{>{\centering\let\newline\\\arraybackslash\hspace{0pt}}m{#1}}
\newcolumntype{R}[1]{>{\raggedleft\let\newline\\\arraybackslash\hspace{0pt}}m{#1}}
\usepackage{booktabs}

\begin{document}
\title{
CEmb-SAM: Segment Anything Model with Condition Embedding for Joint Learning from Heterogeneous Datasets
}
\titlerunning{CEmb-SAM: Segment Anything Model with Condition Embedding}
%
%


\IfStrEq{underreview}{\mystatus}
{


\author{Anonymous\inst{1}} 
\authorrunning{Anonymous et al.}
%
\institute{Anonymous Organization}
}
{
\author{Dongik Shin\inst{1} 
\and
Beomsuk Kim, M.D.\inst{2,3}
\and
Seungjun Baek\inst{1}\thanks{Corresponding author.}
}

\authorrunning{Dong Ik Shin et al.}

\institute{Department of Computer Science, Korea University, Seoul, Republic of Korea\\
\email{\{sdimivy014, sjbaek\}@korea.ac.kr}\\
\and
Department of Physical and Rehabilitation Medicine, Chung-Ang University College of Medicine, Seoul, Republic of Korea\\
\email{grit@cau.ac.kr}\\
\and
Chung-Ang University Gwang-Myeong Hospital, Gwangmyeong-si, Republic of Korea}
}

%

\maketitle              
\input{1.abstract}
\input{2.introduction}

\input{3.method}

\input{4.experiments}
\input{5.conclusion}

\IfStrEq{underreview}{\mystatus}
{
}
{
\subsubsection{Acknowledgements}
This work was supported by the Korea Medical Device Development Fund grant funded by the Korea Government (the Ministry of Science and ICT, the Ministry of Trade, Industry and Energy, the Ministry of Health \& Welfare, the Ministry of Food and Drug Safety) (Project Number: 1711195279, RS-2021-KD000009); the National Research Foundation of Korea (NRF) Grant through the Ministry of Science and ICT (MSIT), Korea Government, under Grant 2022R1A5A1027646; the National Research Foundation of Korea(NRF) grant funded by the Korea government (MSIT) (No.2021R1A2C1007215); the MSIT(Ministry of Science and ICT), Korea, under the ICT Creative Consilience program(IITP-2023-2020-0-01819) supervised by the IITP(Institute for Information \& communications Technology Planning \& Evaluation)
}

%
%
%
\bibliographystyle{splncs04}
\bibliography{ref}
%






\end{document}

%% file: 1.abstract.tex
\begin{abstract}

Automated segmentation of ultrasound images can assist medical experts with diagnostic and therapeutic procedures. Although using the common modality of ultrasound, one typically needs separate datasets in order to segment, for example, different anatomical structures or lesions with different levels of malignancy. In this paper, we consider the problem of jointly learning from heterogeneous datasets so that the model can improve generalization abilities by leveraging the inherent variability among datasets. 
We merge the heterogeneous datasets into one dataset and refer to each component dataset as a subgroup. We propose to train a single segmentation model so that the model can adapt to each sub-group.
For robust segmentation, we leverage recently proposed \textit{Segment Anything model} (SAM) in order to  incorporate sub-group information into the model. We propose SAM with Condition Embedding block (CEmb-SAM) which encodes sub-group conditions and combines them with image embeddings from SAM. The conditional embedding block effectively adapts SAM to each image sub-group by incorporating dataset properties through learnable parameters for normalization. Experiments show that CEmb-SAM outperforms the baseline methods on ultrasound image segmentation for peripheral nerves and breast cancer. The experiments highlight the effectiveness of Cemb-SAM in learning from heterogeneous datasets in medical image segmentation tasks. 

\keywords{Breast Ultrasound \and Nerve Ultrasound \and Segmentation. \and Segment Anything Model}
\end{abstract}

%% file: 2.introduction.tex
\section{Introduction}
\label{sec:intro}
Image segmentation is an important task in medical ultrasound imaging. For example, peripheral nerves are often detected and screened by ultrasound, which has become a convention modality for computer-aided diagnosis (CAD)~\cite{noble2006ultrasound}. As entrapment neuropathies are considered to be accurately screened and diagnosed by ultrasound~\cite{beekman2003sonography,cartwright2013neuromuscular,walker2018indications}, the segmentation of peripheral nerves helps experts identify anatomic structures, measure nerve parameters and provide real-time guidance for therapeutic purposes. In addition, Breast ultrasound images (BUSI) can guide experts to localize and characterize breast tumors, which is also one of the key procedures in CAD~\cite{xian2018automatic}. 

The advancements in deep learning enable an automatic segmentation of ultrasound images, though they still require large, high-quality datasets. 
The scarcity of the labeled data motivated several studies to propose learning from limited supervision, such as transfer learning~\cite{shin2016deep}, supervised domain adaptation~\cite{motiian2017unified,redko2020survey} and unsupervised domain adaptation~\cite{kamnitsas2017unsupervised,dong2018unsupervised,mahmood2018unsupervised}. In practice, separate datasets are needed to train a model to segment different anatomical structures or lesions with different levels of malignancy. For example, peripheral nerves can be detected and identified across different human anatomic structures, such as peroneal (located below the knee) and ulnar (located inside the
elbow) nerves. Typically, the annotated datasets for peroneal and ulnar nerves are separately constructed, and models are separately trained. However, since the models perform a similar task, i.e., segmenting nerve structures from ultrasound images, one may use a \emph{single} model to be jointly trained with peroneal and ulnar nerves in order to leverage the variability in heterogeneous datasets and improve generalization abilities. A similar argument can be applied to breast ultrasound. A breast tumor is categorized into two types, benign and malignant, and we examine the effectiveness of a single model handling the segmentation of both types of lesions. While a simple approach would be incorporating multiple datasets for training, the characteristics of imaging vary among datasets, and it is challenging to train models which deal with distribution shift and generalize well for the entire heterogeneous datasets~\cite{davatzikos2019machine,zhou2021review,wang2022embracing}.

In this paper, we consider methods to train a single model with heterogeneous datasets jointly. We combine the heterogeneous datasets into one dataset and call each component dataset as a \emph{subgroup}. We consider a model which can adapt to domain shifts among sub-groups and improve segmentation performances. We leverage recently proposed \textit{Segment Anything model} (SAM) which has shown great success in natural image segmentation~\cite{kirillov2023segment}. 
However, several studies have shown that SAM could fail on medical image segmentation tasks~\cite{deng2023segment,hu2023sam,he2023accuracy,zhou2023can,ma2023segment}.
We adapt SAM to distribution shifts across sub-groups using a novel method for  condition embedding, which is called SAM with Condition Embedding block (CEmb-SAM). In CEmb-SAM, we encode sub-group conditions and combine them with image embeddings. 
Through experiments, we show that the sub-group conditioning guides SAM to adapt to each sub-group effectively. Experiments demonstrate that, compared with SAM~\cite{kirillov2023segment} and MedSAM~\cite{ma2023segment}, CEmb-SAM shows consistent improvements in the segmentation tasks for both peripheral nerves and breast lesions. Our main contributions are as follows:

\begin{itemize}
    \item We propose CEmb-SAM, which jointly trains a model over heterogeneous datasets leveraging \textit{Segment Anything model} for robust segmentation performances.
    
    \item We propose a conditional embedding module to combine sub-group representations with image embeddings, which effectively adapts the Segment Anything Model to sub-group conditions.
    
    \item Experiments on the peripheral nerve and the breast cancer datasets demonstrate that CEmb-SAM significantly outperforms the baseline models.
    
\end{itemize}

%% file: 3.method.tex
\section{Method}
\label{sec:method}

\begin{figure}[t!]
\begin{center}
\includegraphics[width=0.9\linewidth]{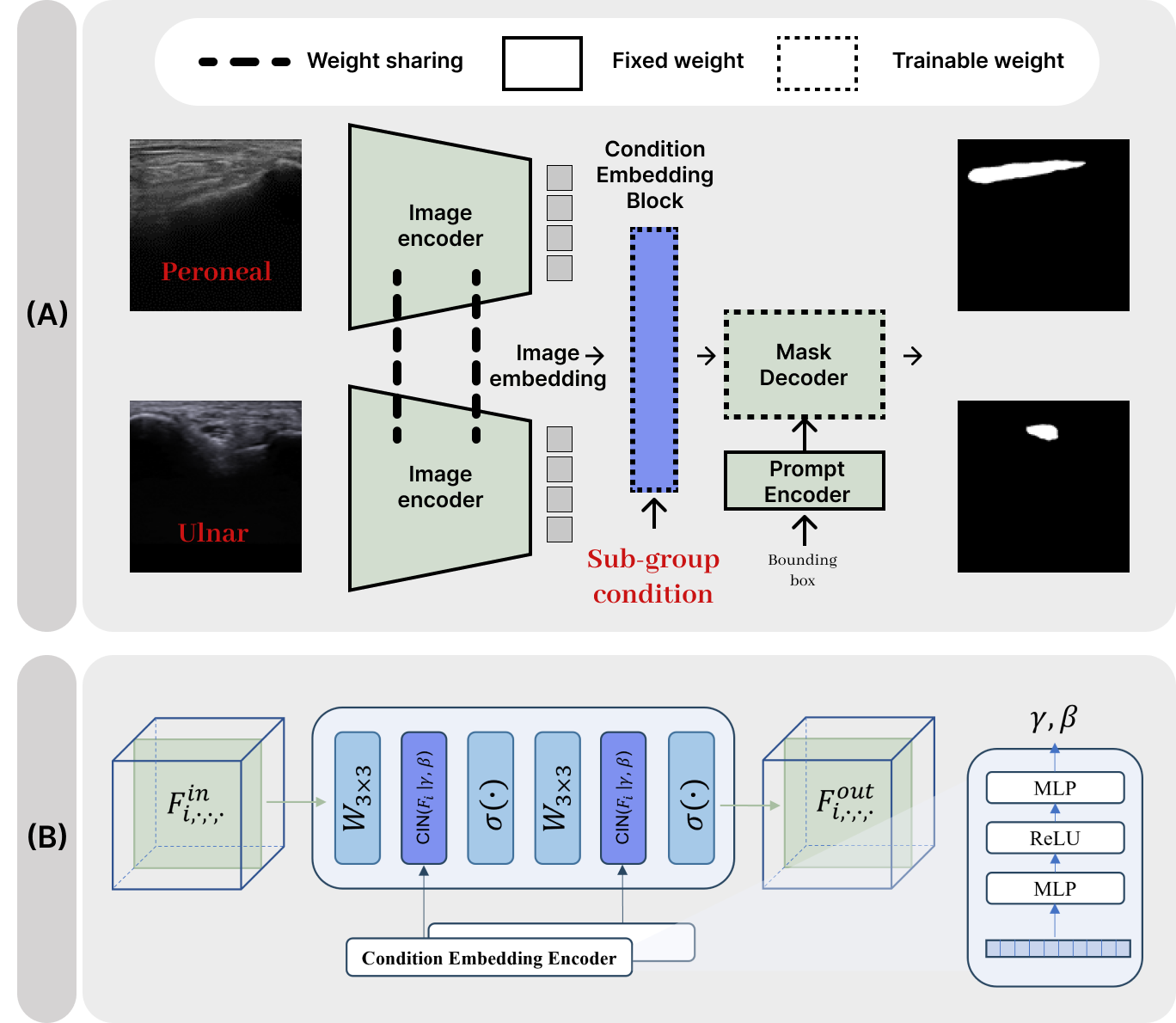}
\end{center}
\vspace{-5mm}
\caption{ (A) CEmb-SAM: \textit{Segment Anything model} with Condition Embedding block. Input images come from heterogeneous datasets, i.e., the datasets of peroneal and ulnar nerves, and the model is jointly trained to segment both types of nerves. The sub-group condition is fed into Condition Embedding block and encoded into sub-group representations. Next, the image embeddings are combined with sub-group representations. The image and prompt encoders are frozen during the fine-tuning of Condition Embedding block and mask decoder. (B) Detailed description of Condition Embedding Block. The sub-group condition is encoded into learnable parameters $\gamma$ and $\beta$, and the input feature $F^{\tiny \textnormal{in}}$ is scaled and shifted using those parameters.}
\label{fig:fig1}
\end{figure}

 The training dataset is a mixture of $m$ heterogeneous datasets or sub-groups. The training dataset with $m$ mutually exclusive sub-groups $\mathcal{D} = \mathbf{g}_{1}\cup \mathbf{g}_{2}\cup \dots \cup \mathbf{g}_{m}$ consists of N samples $\mathcal{D}=\{ (x_{i}, y_{i}, y_{i}^{a})_{i=1}^{N}\}$ where $x_{i}$ is an input image, $y_{i}$ is a corresponding ground-truth mask.
 The sub-group condition $y_{i}^{a} \in \{0, \dots, m-1\}$ represents the index of the sub-group the data belongs to.
 The peripheral nerve dataset consists of seven sub-groups, six different regions at the peroneal nerve (located below the knee) and a region at the ulnar nerve (located inside the elbow). The BUSI dataset consists of three sub-groups: benign, malignant, and normal. The detailed description and sub-group indices and variables are shown in Table~\ref{table:Wcode}.

\subsection{Fine-tuning SAM with Sub-group Condition}
 SAM architecture consists of three components: image encoder, prompt encoder, and mask decoder. Image encoder uses a vision transformer-based architecture~\cite{dosovitskiy2020image} to extract image embeddings. Prompt encoder utilizes user interactions, and mask decoder generates segmentation results based on the image embeddings, prompt embeddings, and its output token~\cite{kirillov2023segment}. We propose to combine sub-group representations with image embeddings from the image encoder using the proposed Condition Embedding block (CEmb). The proposed method, SAM with condition embedding block (CEmb-SAM), uses a pre-trained SAM (ViT-B) model as the image encoder and the prompt encoder. For the peripheral nerve dataset, we fine-tune the mask decoder and CEmb with seven sub-groups. Likewise, we fine-tune the mask decoder on the breast cancer dataset with three sub-groups. The overall framework of the proposed model is illustrated in Fig.~\ref{fig:fig1}. 

\subsection{Condition Embedding Block}
We modified the conditional instance normalization (CIN)~\cite{dumoulin2016learned} to combine sub-group representations and image embeddings. Learnable parameters ${W}_{\gamma}, {W}_{\beta} \in \mathbb{R}^{ {C}\times {m}}$ where ${m}$ is the number of sub-groups of the datasets, and ${C}$ is the number of the output feature maps.
A sub-group condition $y^{a}$ is converted to one-hot vectors, $x^{a}_{\gamma}$ and $x^{a}_{\beta}$ which are fed into Condition Embedding encoder and transformed into sub-group representation parameters $\gamma$ and $\beta$ using two fully connected layers (FCNs). Specifically,

\input{table_wcode}

\begin{equation}
    \gamma = {W}_{2}\cdot\sigma({W}_{1}\cdot {W}_{\gamma}\cdot x^{a}_{\gamma}),\;\beta = {W}_{2}\cdot\sigma({W}_{1}\cdot {W}_{\beta}\cdot x^{a}_{\beta})
\end{equation}

\noindent where ${W}_{1},\,{W}_{2} \in \mathbb{R}^{{C}\times{C}}$ are FCN weights, and $\sigma(\cdot)$ represents ReLU activation function. 

The image embedding $x$ is transformed into the final representation $z$ using the condition embedding as follows. The image embedding is normalized with mini-batch $\mathcal{B}=\{x_{i}, y_{i}^{a}\}_{i = 1}^{N_{n}}$ of $N_{n}$ examples as follows:

\begin{equation}
    \textnormal{CIN}(x_{i}\vert\gamma,\beta) = \gamma \frac{x_{i}-\textnormal{E}[x_{i}]}{\sqrt{\textnormal{Var}[x_{i}]}+\epsilon} + \beta
\end{equation}

\noindent where $\textnormal{E}[x_{i}]$ and $\textnormal{Var}[x_{i}]$ are the instance mean and variance, and $\gamma$ and $\beta$ are given by Condition Embedding encoder. The proposed CEmb consists of two independent consecutive CIN layers with convolutional layers given by:

\begin{equation}
    {F}^{{\tiny \textnormal{mid}}} = \sigma (\textnormal{CIN}(W_{3\times 3}\cdot x_{i}\vert \gamma_{1}, \beta_{1} ))
\end{equation}
\begin{equation}
    z = \sigma (\textnormal{CIN}(W_{3\times 3}\cdot {F}^{{\tiny \textnormal{mid}}}\vert \gamma_{2}, \beta_{2}))
\end{equation}

\noindent where ${F} \in \mathbb{R}^{c\times h\times w}$ represents an intermediate feature map, $\textnormal{W}_{3\times 3}$ denotes convolution kernel size with $3\times 3$. Fig.~\ref{fig:fig1} (B) illustrates the Condition Embedding block.

%% file: table_wcode.tex
\begin{table}[t]
\caption{Summary of the predefined sub-group conditions of peripheral nerve and BUSI datasets. FH: fibular head, FN: fibular neuropathy. $\textnormal{FN}+\alpha$ represents the measured site is $\alpha$ cm away from the fibular head. $m$ represents the total number of sub-groups.}
\vspace{2mm}
\centering
\begin{tabular}{C{1cm}C{1.5cm}C{2cm}C{1cm}C{1cm}C{1.5cm}C{2cm}C{1cm}}
\toprule
Study & Region & Sub-group & $m=7$ & Study & Region & Sub-group & $m=3$\\
\cmidrule(rl){1-4}
\cmidrule(rl){5-8}

\multirow{7}{*}{Nerve} & \multirow{6}{*}{Peroneal} & FH & 0 &  &  &  & \\
 &  & FN & 1 &  &  & \multirow{2}{*}{Benign} & \multirow{2}{*}{0} \\
 &  & FN+1 & 2 & \multirow{3}{*}{BUSI} & \multirow{3}{*}{Breast} &  & \\
 &  & FN+2 & 3 &  &  & Malignant & 1 \\
 &  & FN+3 & 4 &  &  & \multirow{2}{*}{Normal} & \multirow{2}{*}{2} \\
 &  & FN+4 & 5 &  &  &  & \\
 \cmidrule(rl){2-4}
 & Ulnar & Ulnar & 6 &  &  &  & \\
\bottomrule
\end{tabular}
\label{table:Wcode}
\end{table}

%% file: 4.experiments.tex
\section{Experiments}
\label{sec:result}

\input{table_data}
\input{table_result}
\input{images/results/fig_result}

\subsection{Dataset Description}
We evaluate our method on two datasets: (\romannumeral 1) a public benchmark dataset, Breast Ultrasound images (BUSI)~\cite{al2020dataset}; (\romannumeral 2) the peripheral nerve ultrasound images collected in our institution.  
Ultrasound images in the public BUSI dataset are measured from an identical site. The dataset is categorized into three sub-groups: benign, malignant, and normal. The shape of a breast lesion varies according to its type. The benign lesion possesses a relatively round and convex shape. On the other hand, the malignant lesion possesses a rough and uneven spherical shape. The BUSI dataset consists of 780 images. The average image size of the dataset is $500 \times 500$ pixels. 
 
\IfStrEq{underreview}{\mystatus}
{
The peripheral nerve dataset was created at the **** institution.
}
{
The peripheral nerve dataset was created at the Department of Physical Medicine and Rehabilitation, Korea University Guro Hospital.
}
The dataset consists of ultrasound images of two different anatomical structures, the peroneal nerve and the ulnar nerve. The peroneal nerve, on the outer side of the calf of the leg, contains 410 images with an average size of $494 \times 441$ pixels. The peroneal nerve images are collected from six different anatomical structures where the nerve stem comes from the adjacent fibular head. FH represents the fibular head, and FN represents fibular neuropathy. FN+$\alpha$ represents that the measured site is $\alpha$ cm away from the fibular head. The ulnar nerve is located along the inner side of the arm and passing close to the surface of the skin near the elbow. The ulnar nerve dataset contains 1234 images with an average size of $477 \times 435$ pixels. Table~\ref{table:Datasets} describes the sample distribution of datasets. 

\IfStrEq{underreview}{\mystatus}
{
This study was approved by the  Institutional Review Board of ***** (IRB No. *****).
}
{
This study was approved by the Institutional Review Board at Korea University (IRB number: 2020AN0410).
}

\subsection{Experimental Setup}
Each dataset was randomly split at a ratio of 80:20 for training and testing. Each training set was also randomly split into 80:20 for training and validation. SAM comes with three segmentation modes: segmenting everything in a fully automatic way, bounding box mode, and point mode. However, in the case of applying SAM for medical image segmentation, it seems that the segment everything mode is prone to erroneous region partitions. The point-based mode empirically requires multiple iterations of prediction correction. The bounding box-based mode can clearly specify the ROI and obtain good segmentation results without multiple trials and errors~\cite{ma2023segment}. Therefore, we choose the bounding box prompts as input to the prompt encoder for SAM, MedSAM, and CEmb-SAM. In the training phase, the bounding box coordinates were generated from the ground-truth targets with a random perturbation of 0-10 pixels.

The input image's intensity values were normalized using Min-Max normalization~\cite{patro2015normalization} and resized to $3\times256\times256$. We used the pre-trained SAM (ViT-B) model as an image encoder. An unweighted sum between Dice loss and cross-entropy loss is used as the loss function~\cite{isensee2021nnu,ma2021loss}. Adam optimizer~\cite{kingma2014adam} was chosen to train our proposed method and baseline models using NVIDIA RTX 3090 GPUs. The initial learning rate of our model is 3e-4. 

\subsection{Results}
 To evaluate the effectiveness of our method, we compare CEmb-SAM with the U-net~\cite{ronneberger2015u}, SAM~\cite{kirillov2023segment}, and MedSAM~\cite{ma2023segment}. The U-net is trained from scratch on BUSI and peripheral nerve datasets, respectively. The SAM is used with the bounding box mode. The pre-trained SAM (ViT-B) weights are used as image encoder and prompt encoder. During inference, the bounding box coordinates are used as the input to the prompt encoder. Likewise, the pre-trained SAM (ViT-B) weights are used as image encoder and prompt encoder in the MedSAM. The mask decoder of MedSAM is fine-tuned on BUSI and peripheral nerve datasets. CEmb-SAM also uses the pre-trained SAM (ViT-B) model as an image encoder and prompt encoder, and fine-tunes the mask decoder on BUSI and peripheral nerve datasets. During inference, the bounding box coordinates are used as the input to the prompt encoder.

 For the performance metrics, we used the Dice Similarity Coefficient (DSC) and Pixel Accuracy (PA)~\cite{maier2022metrics}. Table~\ref{table:result} shows the quantitative results comparing with CEmb-SAM, MedSAM, SAM (ViT-B), and U-net on both BUSI and peripheral nerve datasets. From Table~\ref{table:result}, we observe that our method achieves the best results on both DSC and PA scores. CEmb-SAM outperformed the baseline methods in terms of the average DSC by 18.61\% in breast, 14.85\% in peroneal, and 14.68\% in ulnar, and in terms of the average PA by 3.26\% in breast, 2.24\% in peroneal and 1.71\% in ulnar.

 Fig~\ref{fig:result} shows the visualization of segmentation results on peripheral nerve dataset and BUSI. The qualitative results show that CEmb-SAM achieves the best segmentation results with fewer missed and false detections in the segmentation of both the breast lesions and peripheral nerves.
 The results demonstrate that CEmb-SAM is more effective and robust in the segmentation through learning from domain shifts caused by heterogeneous datasets.

%% file: table_data.tex
\begin{table}[t]
\caption{Sample distribution of peripheral nerve and BUSI datasets. FH: fibular head, FN: fibular neuropathy. $\textnormal{FN}+\alpha$ represents that the measured site is $\alpha$ cm away from the fibular head.}
\vspace{2mm}
\centering
\begin{tabular}{cccccccc}
\toprule
Dataset & Region & Sub-group & \#of samples & Dataset & Region & Sub-group & \#of samples\\
\cmidrule(rl){1-4}
\cmidrule(rl){5-8}
\cmidrule(rl){1-4}
\cmidrule(rl){5-8}

\multirow{7}{*}{Nerve} & \multirow{6}{*}{Peroneal} & FH & 91 &  &  &  &  \\
 &  & FN & 106 &  &  & \multirow{2}{*}{Benign} & \multirow{2}{*}{437} \\
 &  & FN+1 & 77 &  &  &  &  \\
 &  & FN+2 & 58 & BUSI & Breast & Malignant & 210 \\
 &  & FN+3 & 49 &  &  & \multirow{2}{*}{Normal} & \multirow{2}{*}{133} \\
 &  & FN+4 & 29 &  &  &  &  \\
 & Ulnar & Unknown & 1234 &  &  &  &  \\

\cmidrule(rl){1-4}
\cmidrule(rl){5-8}
Total &  &  & 1644 & Total &  &  & 780 \\
\bottomrule
\end{tabular}
\label{table:Datasets}
\end{table}

%% file: table_result.tex
\begin{table*}[t]
\caption{Performance comparison between U-net, SAM, MedSAM and CEmb-SAM on BUSI and Peripheral nerve datasets.}
\centering
\vspace{2mm}
\begin{tabular}{C{0.9cm}C{1.5cm}C{1cm}C{1cm}C{1.5cm}C{1cm}C{1cm}C{1cm}C{1.5cm}C{1cm}}
\toprule
\multirow{2}{*}{Study} & \multirow{2}{*}{Region} & \multicolumn{4}{c}{DSC (\%)} & \multicolumn{4}{c}{PA (\%)} \\ 
\cmidrule(rl){3-6}
\cmidrule(rl){7-10}
 &  & U-net & SAM & MedSAM & Ours & U-net & SAM & MedSAM & Ours\\
\cmidrule(rl){1-2}
\cmidrule(rl){3-6}
\cmidrule(rl){7-10}
 BUSI & Breast & 64.87 & 61.42 & 85.95 & \textbf{89.35} & 90.72 & 87.19 & 90.89 & \textbf{92.86}\\
\cmidrule(rl){1-2}
\cmidrule(rl){3-6}
\cmidrule(rl){7-10}
 \multirow{2}{*}{Nerve} & Peroneal & 69.91 & 61.72 & 78.87 & \textbf{85.02} & 92.59 & 90.58 & 91.81 & \textbf{93.90}\\
  & Ulnar & 77.04 & 59.56 & 83.98 & \textbf{88.21} & 96.49 & 94.89 & 96.66 & \textbf{97.72}\\
\bottomrule
\end{tabular}

\label{table:result}
\end{table*}

%% file: images/results/fig_result.tex
\begin{figure}[t]
\begin{center}
\includegraphics[width=\columnwidth]{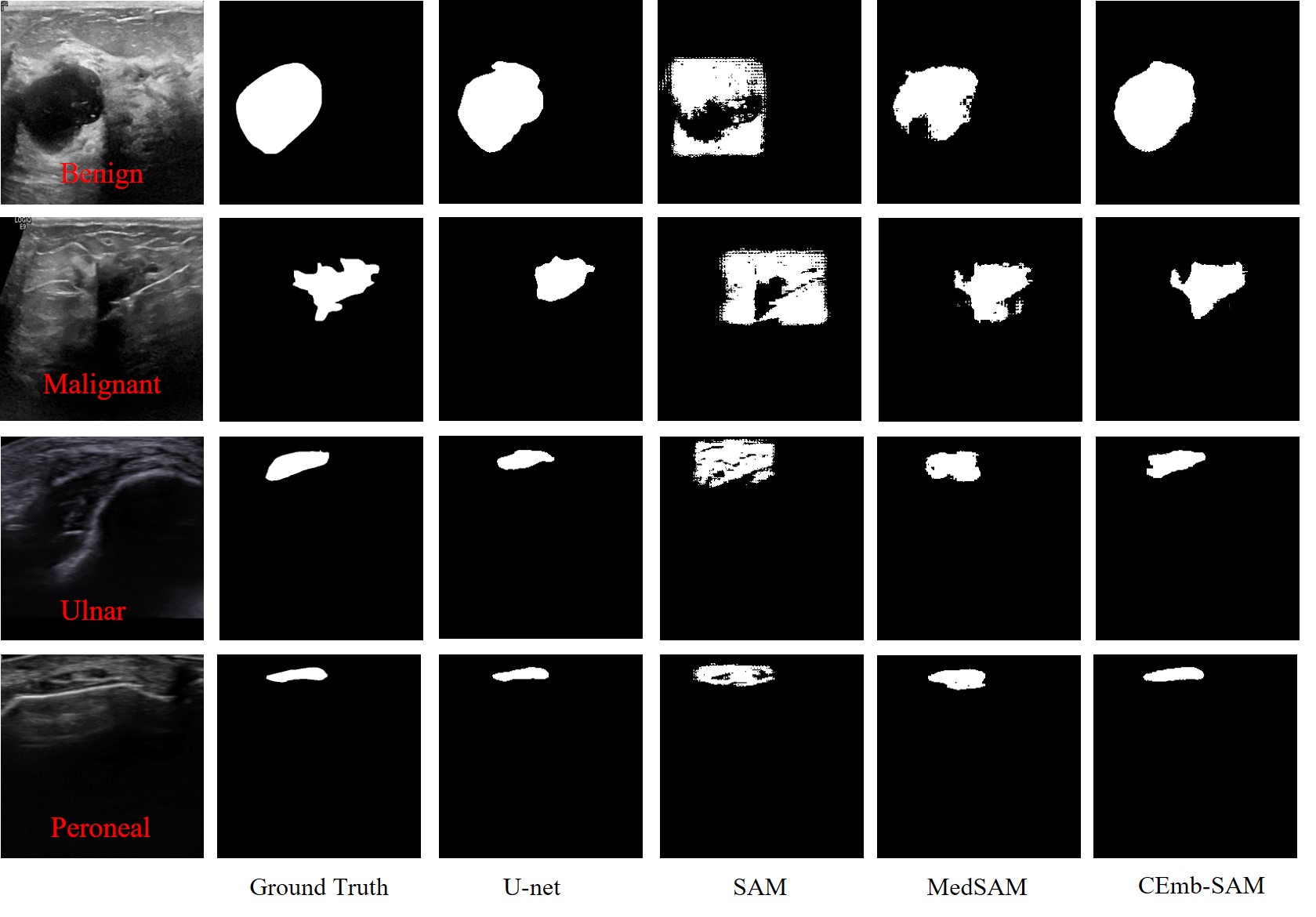}
\end{center}
\vspace{-5mm}
\caption{Segmentation results on BUSI (1st and 2nd rows) and peripheral nerve dataset (3rd and 4th rows).}
\label{fig:result}
\end{figure}

%% file: 5.conclusion.tex
\section{Conclusion}
\label{sec:con}
In this study, we propose CEmb-SAM which adapts the Segment Anything Model to each dataset sub-group for
joint learning from the entire heterogeneous datasets of ultrasound medical images. The proposed module for conditional instance normalization was able to guide the model to effectively combine image embeddings with subgroup conditions for both the BUSI and peripheral nerve datasets. The proposed module helped the model deal with distribution shifts among sub-groups. Experiments showed that CEmb-SAM achieved the highest score in DSC and PA on both the public BUSI dataset and peripheral nerve datasets. As future work, we plan to extend our work for improved domain adaptation in which the model is robust and effective under higher degrees of anatomical heterogeneity among datasets.

